\documentclass[10pt,preprint]{aastex}
\usepackage{amsmath}
\usepackage{graphicx}
\usepackage{epsfig}
\usepackage{hyperref}

\usepackage{natbib}
\usepackage{dcolumn}
\newcolumntype{d}[1]{D{.}{\cdot}{#1} }

\usepackage[normalem]{ulem}
\usepackage[usenames]{color}
\definecolor{mygreen}{rgb}{0.0, 0.5, 0.0}
\definecolor{mylila}{rgb}{0.5, 0.0, 0.5}
\definecolor{myblue}{rgb}{0.0, 0.0, 0.5}

\newcommand{\rpmj}{RPM$_J$}

\title{Target Selection for the SDSS-III MARVELS Survey}
\author{Martin Paegert\altaffilmark{1}, 
Keivan G.\ Stassun\altaffilmark{1, 2}, 
Nathan De Lee\altaffilmark{3, 9}, 
Joshua Pepper\altaffilmark{4}, 
Scott W.\ Fleming\altaffilmark{5, 6, 9}, 
Thirupathi Sivarani\altaffilmark{7, 9}, 
Suvrath Mahadevan\altaffilmark{10, 11, 9}, 
Claude E. Mack III\altaffilmark{1}, 
Saurav Dhital\altaffilmark{12},
Leslie Hebb\altaffilmark{8},
Jian Ge\altaffilmark{9}
}
\altaffiltext{1}{Vanderbilt University, Nashville, TN 37235, USA}
\altaffiltext{2}{Fisk University, Nashville, TN 37208, USA}
\altaffiltext{3}{Northern Kentucky University, Highland Heights, KY 41099, USA}
\altaffiltext{4}{Lehigh University, Bethlehem, PA 18015, USA}
\altaffiltext{5}{Computer Sciences Corporation, 3700 San Martin Dr, Baltimore, MD, USA 21218}
\altaffiltext{6}{Space Telescope Science Institute, 3700 San Martin Dr, Baltimore, MD, USA 21218}
\altaffiltext{7}{Indian Institute of Astrophysics, Koramangala, Bangalore-560034, India}
\altaffiltext{8}{Hobart \& William Smith Colleges, Ithaca, NY, USA}
\altaffiltext{9}{University of Florida, 2014 Turlington Hall, Gainesville FL 32611, USA}
\altaffiltext{10}{Department of Astronomy and Astrophysics, The Pennsylvania State University, University Park, PA 16802, USA }
\altaffiltext{11}{Center for Exoplanets and Habitable Worlds, The Pennsylvania State University, University Park, PA 16802, USA}
\altaffiltext{12}{Department of Physical Sciences, Embry-Riddle Aeronautical University, 600 South Clyde Morris Blvd., Daytona Beach, FL 32114, USA}

\begin{document}

\begin{abstract}
We present the target selection process for the Multi-object APO Radial Velocity Exoplanets Large-area Survey (MARVELS), which is part of the Sloan Digital Sky Survey (SDSS) III. MARVELS  is a medium-resolution ($R \sim 11000$) multi-fiber spectrograph capable of obtaining radial velocities for 60 objects at a time in order to find brown dwarfs and giant planets. The survey was configured to target dwarf stars with effective temperatures approximately between $4500$ and $6250 \, \mbox{K}$.  For the first 2 years MARVELS relied on low-resolution spectroscopic pre-observations to estimate the effective temperature and $\log(g)$ for candidate stars and then selected suitable dwarf stars from this pool. Ultimately, the pre-observation spectra proved ineffective at filtering out giant stars; many giants were incorrectly classified as dwarfs,
resulting in a giant contamination rate of $\sim$30\% for the first phase of the MARVELS survey.
Thereafter, the survey instead applied a reduced proper motion cut to eliminate giants and used the Infrared Flux Method to estimate effective temperatures, using only extant photometric and proper-motion catalog information. The target selection method introduced here may be useful for other surveys that need to rely on extant catalog data for selection of specific stellar populations. 
\end{abstract}

\textit{Accepted by AJ}

\section{Introduction\label{sec:intro}}
Target selection is a crucial step for most astronomical surveys, one that may have a significant impact on the result even before the first image is taken. This is especially true for exoplanet surveys. A common method is to pre-select stars according to brightness, then derive stellar parameters from reconnaissance observations and compile them into an Input Catalog from which the final set of targets is drawn. The NASA {\it Kepler} mission is one of the most prominent projects following this process.  
However, such pre-observations require telescope time and extensive effort to process and evaluate the reconnaisance data. 
Therefore, for MARVELS
\citep[Multi-Object APO Radial Velocity Exoplanet Large-area Survey][]{ge09}
we opted to devise a technique 
to find the stellar populations suiting the scientific purposes of the survey
using only existing catalog data, thus saving the time and effort that would otherwise go into pre-observations,
and streamlining the target selection process significantly. 
We hope that our method will be useful for future surveys like the upcoming Transiting Exoplanet Survey Satellite (TESS), or indeed any effort to select a particular population of stars from existing catalog data. 

MARVELS is part of the SDSS--III program \citep{sdssiii} and uses a specially built 60-fiber spectrograph to obtain medium-resolution ($R = 11000$) spectra to derive the precision radial velocities needed to find exoplanets and brown dwarfs orbiting main sequence stars. 
The MARVELS instrument itself is described in Ge et al.\ (2015a,b; in preparation), the data reduction pipeline is described in Thomas et al.\ (2015; in preparation), and the final DR12 data release is described in Alam et al.\ (2015; in preparation). In this paper we focus on describing the MARVELS target selection process.

For each target field with a circular field of view of 7 square degrees, 56 stars are selected for observation and assigned to a fixed fiber which then is plugged to a hole in a metal plate placed in the focal plane of  2.5 m SDSS telescope \citep{sdss25m}. 
Four fibers are reserved for guide stars, which are chosen after the science targets are known. The plugs require a minimal distance of $75$ arcsec and thus define the required minimal distance between target stars.  
Between October 2008 and July 2012, MARVELS made 1565 observations of 92 fields collecting multi-epoch data for 5520 stars, more than 90\% of them with enough epochs to be processed through the pipeline and yield sufficient RV observations to search for companions, including stellar companions, brown dwarfs and giant planets.

Due to technical and administrative changes in January 2011---change of fibers, joint observation with the APOGEE SDSS-III survey (the APO Galactic Evolution Experiment, \citet{apogee}, \citet{eisenstein2011})---the observation is divided into two different phases: before and after January 2011, hereafter referred to as ``initial" (Years 1--2) and ``final" (Years 3--4) phases. 

Section~\ref{sec:targetsel} first describes the final target selection process used for fields observed after January 2011. It then describes the initial process that was used prior to January 2011 as well as the lessons learned and why the initial process was abandoned. Section~\ref{sec:results} presents a summary of the properties of the targets observed. We conclude with a brief summary in Section~\ref{sec:summary}.

\section{Target Selection Methods\label{sec:targetsel}}

MARVELS observed 5520 stars over four years, observing 54 science targets per field at a time. MARVELS was designed to achieve a radial velocity precision of $< 30 \; \mbox{m/s}$  for stars as faint as $V = 12$ magnitudes in order to discover brown dwarfs and giant planets of a homogenous sample of targets with only few, well-understood biases. 
Prime targets for MARVELS are FGK dwarfs, limiting the effective temperatures ($T_{\rm eff}$) to $3500 < T_{\rm eff} \le 6250 \; \mbox{K}$. Most giants---defined by $\log(g) < 3.0$ during the initial phase, $\log(g) < 3.5$ during the final phase---are excluded from the survey because many giant stars exhibit pulsation-driven radial velocity variations that dominate the radial velocity signals of orbiting giant planets or brown dwarfs. In addition, the photospheres of red giants can be very extended, up to $\sim$AU scales, precluding companions with orbits up to $\sim$1 yr. At the same time, for the brightest giants MARVELS could achieve a signal-to-noise ratio good enough for detecting intermediate-period ($\gtrsim$1 yr) giant planets and contribute knowledge about planets around evolved stars. For this reason, 6 fibers (or $10 \%$) per field were reserved for the brightest giants.

\subsection{\label{sectfinal} Final target selection process}

This section describes the final process in effect between January 2011 and the end of observations in July 2012. During this time MARVELS shared the SDSS telescope with APOGEE. Both projects had to observe the same field and coordinate field and target selection, because the fibers for both spectrographs were plugged to the same metal plate.  Therefore, each individual star could only be observed by one of the two surveys. The field names, center coordinates, and number of observations for the joint fields are listed in Table \ref{y34obs}.

\subsubsection{\label{sectinputcat} Basic input catalog construction}

The basic catalog for the final MARVELS target selection is a modified Guide Star Catalog version 2.3 (GSC 2.3) \citep{gsc23}. Although the GSC catalog includes most of the key information required for MARVELS target selection, the published catalog does not include proper motions. The GSC authors state that for the northern sky, proper motion errors are of the order of $6 - 8 \; \mbox{mas/yr}$,  while errors for the southern sky are much larger, and they prefer to release the catalog without any proper motions until the systematic behind this discrepancy is better understood. The proper motion errors in the northern sky are small enough for the purposes of MARVELS, and we obtained a version of GSC 2.3 that includes proper motions for the northern sky. This catalog has been cross-matched with 2MASS \citep{tmass} and is hereafter referred to as ``modified GSC 2.3". 

Some of the fields observed between January 2011 and July 2012 do not have proper motions in the modified GSC2.3. In this case we used Nomad \citep{nomad} or UCAC3 \citep{ucac3}. While Nomad includes 2MASS $JHK$ photometry from a previous match, it does not include a 2MASS designation or identifier, so we re-matched Nomad to 2MASS allowing for a positional error of $1.41$ arcsec radius in J2000 coordinates and a rounding error for $JHK$ of 0.002 mag in each band. Stars without a matching 2MASS entry were rejected. 

We repeated the catalog matching for every target field and then applied a series of steps as given in Table \ref{final_selsteps} to select optimal target stars. The final process includes a $J$-band reduced proper motion (\rpmj) cut to filter out giants and an estimate of the effective temperature using the infrared flux method (IRFM, \citet{casagrande10}). These steps replaced pre-observations of 1000 stars per field, which measured $\log(g)$ and $T_{\rm eff}$ from low-resolution spectra and subsequently selecting the 100 best suited target stars, as was done in the initial phase (see Section~\ref{sec:initial}). The final process allowed us to select the 100 stars per field to be observed with the MARVELS spectrograph from existing catalog data only and thus greatly streamlined the process. The initial limit of 1000 candidate stars in step 8 of Table \ref{final_selsteps} was not strictly necessary but was a vestige from the initial target selection process (Section \ref{sec:initial}) and was retained for consistency.

The reason for selecting 100 stars for 56 fibers is that, during the initial phase, in many cases two fields were drilled on the same metal plate (double drilling), and as a result it was possible for stars from field A to collide with stars from field B. 
The final target selection for each field was done by the plate-drilling team,
thus providing that team with a generous ``reserve" of extra target stars took care of potential conflicts between fields on double drilled plates. During this process each star was assigned to a certain physical fiber, and once this assignment was made the same fiber was used for all future observations of the target.

\begin{table}[htb]
    \begin{tabular}{|c|p{7.5cm}|p{7.0cm}|}
        \hline
        Step & Criterion & Reason \\ \hline
          1  & Keep stars with $7.6 \le V \le 13.0$ & Include only stars in MARVELS magnitude limits \\
          2  & Keep stars with $J - K_{S} \ge 0.29$    & Exclude stars that are clearly too hot \\
          3  & Keep stars with known proper motions & Allows exact positioning of fibers and permits use of reduced proper motion cut \\
          4  & Ensure that positional coordinates, after correcting for proper motion, indicate that the star is in the field for at least 2 years from projected start of observations & Exclude stars that might wander off-plate \\
          5  & If two stars are closer than $75{''}$, keep the brighter star & Prefer bright stars for good SNR\\
  
          6  & Closest star with $V < 9$ must be more than $5{''}$ away & Prevent flux contamination of target star \\
          7  & Exclude star if too close to APOGEE targets & Prevent fiber collision between APOGEE and MARVELS \\
          8  & Limit the results to the 1000 brightest stars in V & Build large enough pool for subsequent steps \\
          9  & Apply reduced proper motion cut to filter out all but the 6 brightest giants & Only 6 giants wanted \\
         
         10  & Exclude hot stars ($T_{\rm eff} > 6250 \mbox{K}$) according to Infrared Flux Method & Exclude hot stars \\
         11  & Limit F stars (those with $5800 \mbox{K} \le T_{\rm eff} \le 6250 \mbox{K}$) to 40\% of all MARVELS targets in the field & Guarantee 50\% GK stars \\
         12  & Limit the total number to 100 per field & 60 plugged, 40 as "reserve" in case of collisions \\
         13  & Check the 6 selected giants in Simbad & Verify for $T_{\rm eff}$, exclude close binaries and known variables \\
       \hline
    \end{tabular}
    \centering
    \parbox{16cm}{
        \caption[final selection steps]{\label{final_selsteps} Target selection steps during the final phase. $T_{\rm eff}$ from Casagrande relations.}}
\end{table}

\subsubsection{\label{sect:rpmj} Reduced Proper Motion Cut}

To get an estimate of $\log(g)$, we first compute the reduced proper motion in $J$ (\rpmj). With $\mu_{r}$, $\mu_{d}$ as proper motion in right ascension and declination in arc-seconds per year and $d$ as declination, we compute
\begin{align}
  \mu  &= \sqrt{(\cos d * \mu_{\rm r})^2 + \mu_{\rm d}^2}         \\
  {\rm RPM_{J}} &= J + 5 \log(\mu)   \label{eq:2}
\end{align}
and then apply an empirical \rpmj\ cut described in \cite{cameron07}:
\begin{equation}
y = -58 + 313.42 (J - H) - 583.6 (J - H)^2 + 473.18 (J - H)^3 - 141.25 (J - H)^4   \label{eq:3}
\end{equation}
Stars with $y \ge$ \rpmj\ are regarded as \rpmj-dwarfs, stars with $y < {\rm RPM_{J}}$ as \rpmj-giants.

\begin{figure}[htb]
\epsfig{file=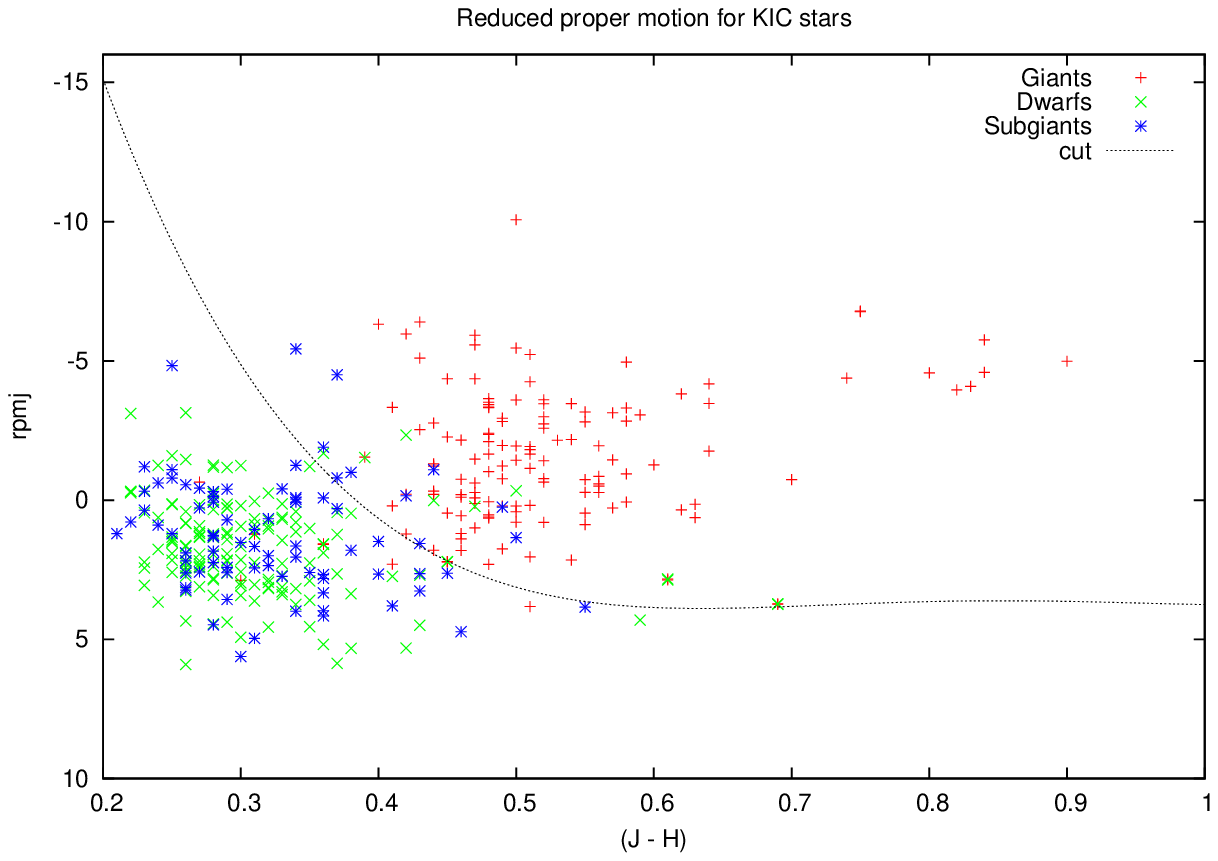, scale=0.65}
\epsfig{file=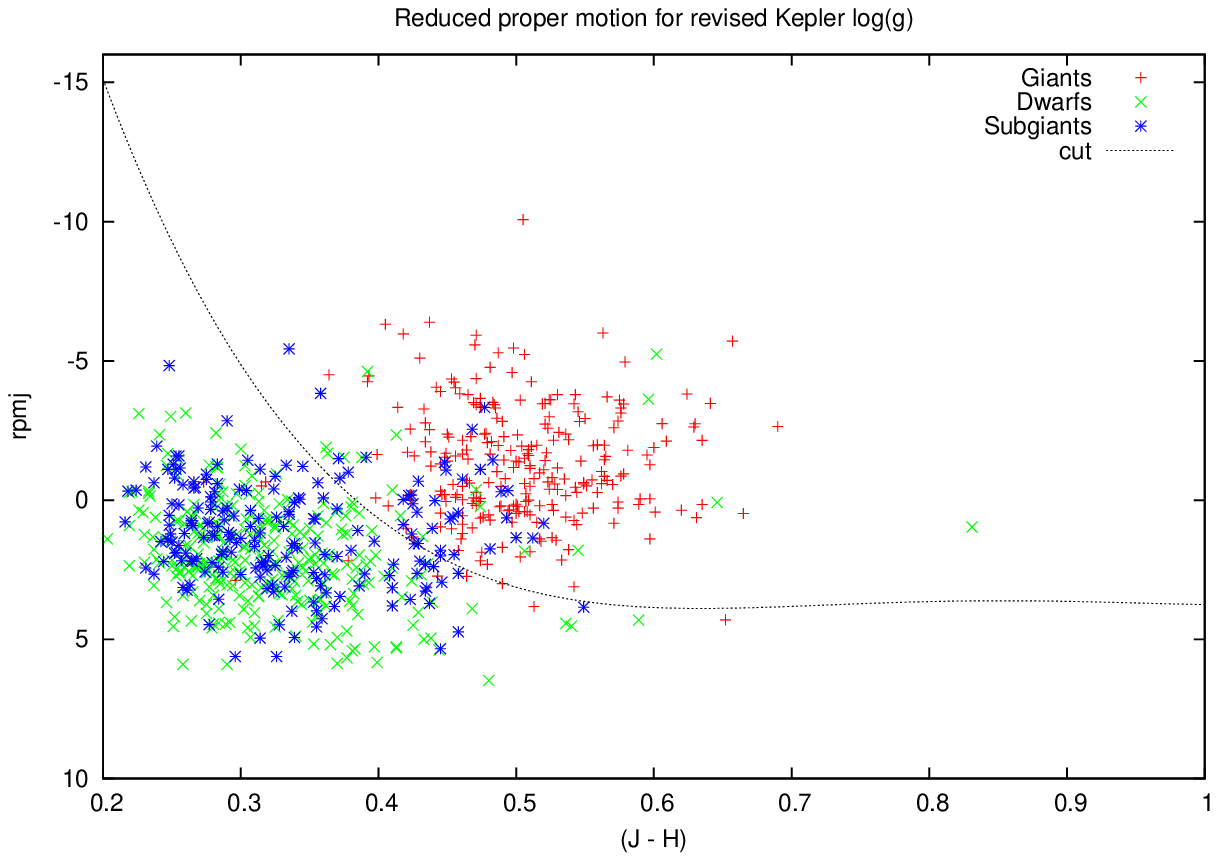, scale=0.65}
\centering
\parbox{16cm}{
\caption[Reduced proper motion cuts] {\label{gilliland_kepler} Reduced proper motion for KIC stars with SDSS spectra using equations (\ref{eq:2}). Red crosses: giants, blue stars: subgiants, green x: dwarfs, solid line: cut according to Equation \ref{eq:3}. Note that the y-axis is inverted. Left panel: log(g) values from original KIC, right panel: revised $\log(g)$ values from NASA Exoplanet Archive; stars with $J - H > 0.7$ are evolved giants that were only observed by Kepler in Q0 and not observed in later quarters. }}
\end{figure}

Using proper motions from the modified GSC we first tested this cut by applying it to the 458 stars in the Kepler Input Catalog \citep[KIC;][]{kic} that also have been observed by MARVELS. Note that the KIC was principally intended for broadly discriminating highly evolved giants from dwarfs, and thus the surface gravities in particular have relatively large uncertainties \citep[e.g.,][]{casagrande14}.
In the time since the MARVELS target selection strategy was implemented and tested, the NASA Exoplanet Archive has released updated stellar characteristics---namely $\log(g)$ and $T_{\rm eff}$---for stars observed by Kepler in quarters 1 to 16. While we used the original KIC values for target selection, for the comparisons shown in this paper we also re-ran the analysis with revised values from the NASA Exoplanet Archive. As we show below, the results using the updated KIC are not substantially different from that used in our actual target selection process.

We define giants as having $\log(g) \le 3.5$, dwarfs having $\log(g) \ge 4.1$, and subgiants as those with $\log(g)$ between these values. Figure \ref{gilliland_kepler} shows the \rpmj\ diagram for the MARVELS-Kepler overlap stars (using original KIC values in the left panel, revised values in the right panel). The solid black line marks the border between \rpmj-dwarfs below and \rpmj-giants above the line as defined by equation (\ref{eq:3}). Green symbols above the line are $\log(g)$-dwarfs that are mis-identified as giants by the \rpmj\ cut. Red symbols below the line are in turn $\log(g)$-giants mis-identified as dwarfs. The MARVELS ``region of interest" is $0.3 < (J-H) < 0.54$ and below the \rpmj\ cut, which translates to dwarfs or subgiants of spectral types F9 to K3. 

A few stars on the left panel in Figure \ref{gilliland_kepler} are missing in the right panel---most notably those with $J - H > 0.7$. These are evolved giant stars that were observed by Kepler in quarter 0 only then dropped and are thus not part of the revised values published in the NASA Exoplanet Archive for stars observed by Kepler in quarters 1 to 16. Nonetheless, the right panel with the revised values is more populated and confirms the ability of the \rpmj\ cut to distinguish between dwarfs and giants---although again subgiants are not well discriminated.

Using the original KIC values we find that $6$ stars ($1.7\%$) are false negatives---$\log(g)$ dwarfs according to KIC, but giants according to the \rpmj\ cut. Another $6$ stars $1.7\%$ are false positives---they are $\log(g)$ giants, falsely identified as dwarfs by the \rpmj\ method. 
Most of the sub-giants are below the line and thus in the ``dwarf"-region. 
\citet{cameron07} included only stars with $\log(g) < 3.0$ or $\log(g) > 4.0$, thus excluding subgiants. In our analysis we included the missing $\log(g)$ interval and conclude that the \rpmj\ cut does not seem able to distinguish sub-giants from dwarfs. For MARVELS this is not a problem, because sub-giants are valid target stars, but it should be considered for any future statistics derived from the MARVELS dataset, as subgiants will be included in the ``dwarf" sample. 

\begin{figure}[htb]
\epsfig{file=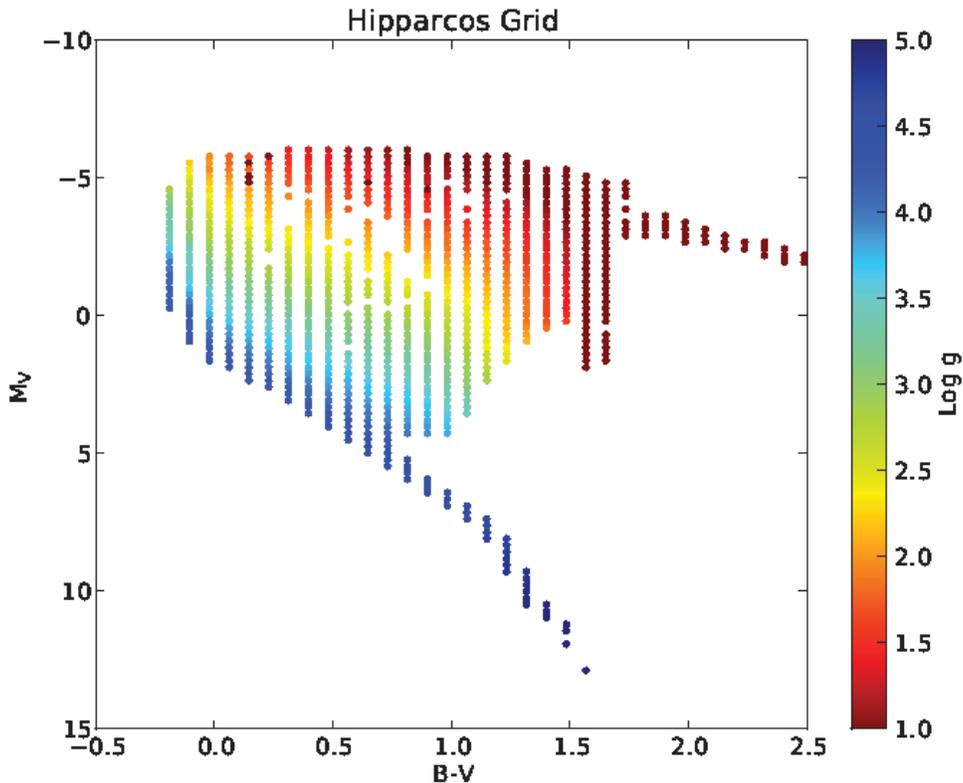, scale=0.75}
\centering
\parbox{16cm}{
\caption[HR Diagram of $\log(g)$ Bins from Isochrones]{\label{isogrid} In this HR diagram, $\log(g)$ from isochrones are plotted. The color of each bin is the median $\log(g)$ of all the isochrones that pass through that color-magnitude bin.}}
\end{figure}

Note that the KIC stars span a magnitude range from $V = 9$ to $11.5$, thus excluding the bright and faint end of the MARVELS magnitude range ($7.6$ to $13.0$). Therefore, one concern is that the good performance of the \rpmj\ cut for the KIC stars might not apply to the full range of MARVELS targets. In addition, when considering stars in other parts of the sky, one would necessarily need to use proper motions from heterogeneous catalog sources, thus potentially introducing systematic errors.
Thus for the MARVELS target stars we checked how much of an influnce a change from the modified GSC to UCAC as source catalog for proper motions would have on our selection. Generally GSC and UCAC are in good agreement, with a mean difference in total proper motion of $\Delta \mu = 0.37 \pm 4.75 \; \mbox{masec/yr}$. Thus on average we do not expect dramatic changes. We also checked the rate of stars switching from giant to dwarf classification according to the \rpmj\ cut when changing from GSC to UCAC as the proper motion source catalog. We found this rate to be $1.75 \%$, thus low enough to not cause concern.

\begin{figure}[htb]
\epsfig{file=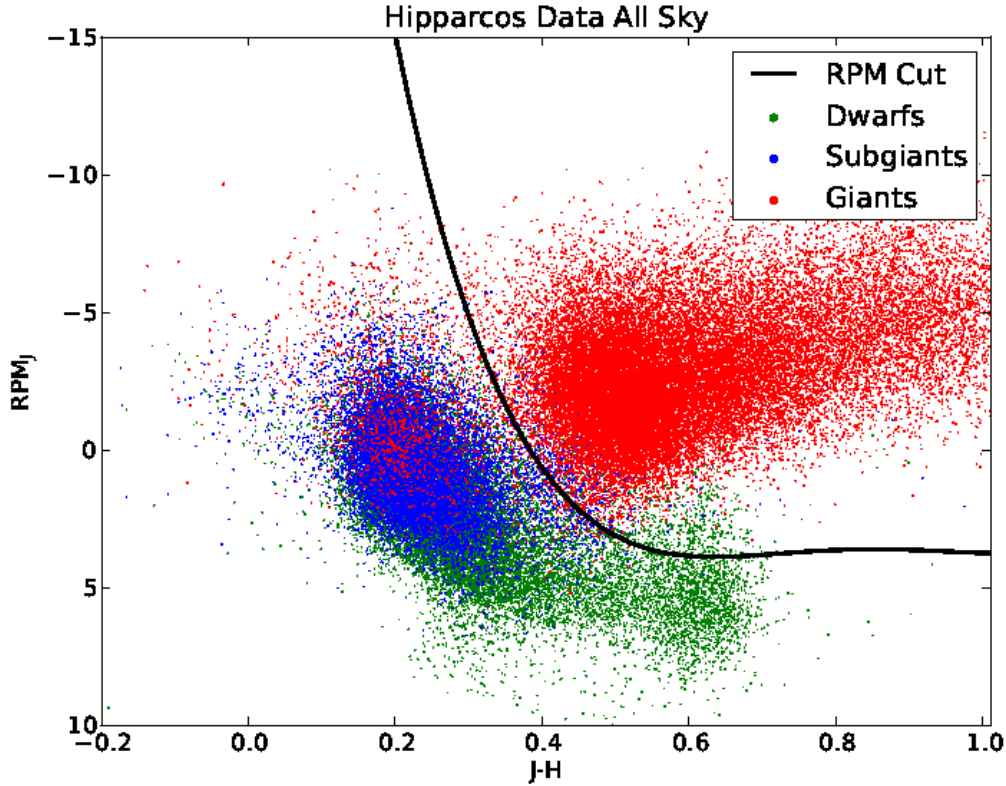, scale=0.75}
\centering
\parbox{16cm}{
\caption[Reduced proper motion for Hipparcos stars] {\label{hiprpmj} Reduced proper motion cut for Hipparcos stars as described in section \ref{sect:rpmj}: red points are giants; blue points are subgiants; green points are dwarfs.}}
\end{figure}

Encouraged by these comparisons with the KIC we next tested the \rpmj\ cut all-sky with the Hipparcos catalog \citep{hip,hip2}. We combined the Hipparcos catalog with isochrones to derive a $\log(g)$ determination based upon location within the HR diagram. We started with the Padova isochrones \citep{marigo08,girardi10} from CMD 2.3  (http://stev.oapd.inaf.it/cmd). The isochrones were chosen to be solar metallicity ($Z = 0.019$) and range to in $\log_{10}(Age) \; \mbox{[yrs]}$ from 6.60  to 10.10 (inclusive) in steps of 0.05. The intrinsic luminosities from the isochrones were transformed into the Johnson-Cousins filters using  \cite{maiz06} and \cite{bessel90}. The HR diagram is a phase space of absolute $V$ magnitude and $B-V$ color. Each isochrone occupies a particular region of this phase space, but they fall in such a way that some regions have multiple isochrones overlapping and some regions have no isochrones. In order to quantify this, the phase space was separated into 40 bins in $B-V$ and 80 bins in the M$_V$. The minimum value, maximum value, and step size in $B-V$ were $-0.23$, 3.12, and 0.0837, respectively. For M$_V$ the minimum value, maximum value, and step size were $-6.14$, 13.00, and 0.239, respectively. The isochrones are made up of a series of three coordinate data points (M$_V$, $B-V$, and $\log(g)$). Each isochrone data point is put into its appropratiate bin in color-magnitude phase space. Once this was done, the median $\log(g)$ of the data points in each color-magnitude bin was assigned as the $\log(g)$ of that bin as shown in Figure \ref{isogrid}. The isochrones do not completely cover the HR diagram, so there are bins that do not have a $\log(g)$ value. The next step is to associate each Hipparcos star with a bin and throw out any stars that do not fall within 0.5 magnitudes (in both M$_V$ or $B-V$) of a bin center that had a $\log(g)$ value. If a Hipparcos star has more than 1 bin within the 0.5 magnitude bin radius, then the star was associated with the closest bin center and was then assigned the $\log(g)$ of that bin. This results in a table of Hipparcos stars with $\log(g)$ that can then be used to test the \rpmj\ method, see Figure \ref{hiprpmj}. Note that our use here of 0.5-mag bins, and of a linear interpolation in log age, are simplified and arbitrary choices, however they suffice for the purposes of the check we seek to perform of the broad performance of the \rpmj\ method to distinguish dwarfs from giants.

To compare the \rpmj\ results with directly measured $\log(g)$ values, we chose the RAVE catalog, data release 2 \citep{rave2} and 3 \citep{rave3}. Both RAVE releases give comparable results, except that the fraction of $\log(g)$ giants which would be classified as \rpmj-dwarfs is nearly three times higher with data release 2 ($15\%$ instead of $5.8\%$). We ascribe this to the improved RAVE pipeline used for data release 3. 

The results for Hipparcos and RAVE DR3 are summarized in Table \ref{hip_rave}. If the \rpmj\ cut is the only information used, $92.5\%$ of stars flagged as giants would be true giants according to the $\log(g)$ value derived for Hipparcos stars. $2.6\%$ would be subgiants and $4.9\%$ would be dwarfs. Furthermore $2.6\%$ of the stars flagged as dwarfs would be giants, $39.9\%$ subgiants and $57.5\%$ dwarfs. 

The results for all stars in RAVE DR3 are similar with a notable shift from sub-giants towards dwarfs and giants. While the ratio of correctly identified giants decreases by $9 \%$, the ratio of correctly identified dwarfs improves by $17 \%$. If we limit the RAVE stars to valid MARVELS targets---those matching the MARVELS magnitude and color cut ($7.6 \le V < 13$, respectively ($J - K_{S} \ge 0.29$)---the results improve slightly, but not significantly, as shown in the bottom part of Table \ref{hip_rave}.

\begin{table}
    \begin{tabular}{|l|c|c|c|}
        \hline
        Hipparcos    & true giants   & true subgiants &  true dwarfs   \\ \hline
        \rpmj\ giants  & 34542 (92.5\%) &   954 (2.6\%)  &  1847 (4.9\%) \\
        \rpmj\ dwarfs  &   889 (2.6\%)  & 13746 (39.9\%) & 19791 (57.5\%) \\
        \hline
        RAVE (all)   &                 &                  &                 \\ \hline
        \rpmj\ giants  & 13028 (84.8\%) &  1304 (8.5\%)  &  1025 (6.7\%) \\
        \rpmj\ dwarfs  &   809 (3.8\%) &  5093 (23.8\%)  & 15464 (72.4\%) \\
        \hline
        RAVE (selected)  &                &                 &                 \\ \hline
        \rpmj\ giants   & 9940 (86.3\%) &  927 (8.1\%)  &   433 (5.6\%) \\
        \rpmj\ dwarfs   &  648 (3.2\%) & 2787 (20.9\%)  & 10136 (75.9\%) \\ \hline
    \end{tabular}
    \centering
    \parbox{8cm}{
        \caption[\rpmj\ classification]{\label{hip_rave} Classification of Hipparcos and RAVE stars by \rpmj\ cut.  The bottom category of RAVE (selected) is restricted to stars in the MARVELS regime for $V$-mag and $J-K$ color.}}
\end{table}

For MARVELS we conclude that using the \rpmj\ cut as the only method for selecting dwarfs will result in a giant contamination rate of about 4\%, which is much better than the rate we experienced from spectroscopic pre-observations (see Section \ref{sec:initial}).
Importantly, however, subgiants comprise a large fraction of the ``dwarf" sample. Therefore while the target selection procedure described above is highly effective at removing evolved red giants, subgiants are unavoidably mixed in with the dwarfs at the level of 20--40\% (see Table \ref{hip_rave}).

\subsubsection{Effective Temperature from Infrared Flux Method}

We compute the effective temperature using color-metallicity-temperature relations based on the Infrared Flux Method (IRFM) as described in \citet{casagrande10}. For stars with $0.78 <= V - K <= 3.15$ and $-5.0 <= [\rm Fe/ \rm H] <= 0.4$, and defining $x = V - K$, and $T_{\rm eff} = 5040.0 / \theta_{\rm eff}$, \citet{casagrande10} gives the relation
\begin{equation}
\theta_{\rm eff} = 0.5057 + 0.2600 x - 0.0146 x^2  - 0.0131 x [\rm Fe/ \rm H] + 0.0288 [\rm Fe / \rm H] + 0.0016 [\rm Fe / \rm H]^2   \label{eq:4}
\end{equation}

Using instead the $J - K$ colors, for stars with $0.07 \le J - K_{S} \le 0.80$ and the same metallicity restrictions, and now defining $x = J - K_{S}$:
\begin{equation}
\theta_{\rm eff} = 0.6393 + 0.6104 x + 0.0920 x^2 - 0.0330 x [\rm Fe / \rm H] + 0.0291 [\rm Fe / \rm H] + 0.0020 [\rm Fe / \rm H]^2    \label{eq:5}
\end{equation}

Without a measured value for $[\rm Fe/ \rm H]$, we assume solar metallicity for all our target stars. The additional error induced by this assumption does not exceed $80 \; \mbox{K}$ at the extreme ends of the color range and for $[\rm Fe/ \rm H] = \pm 0.4$. Comparing $T_{\rm eff}$ from the Casagrande relations with the original $T_{\rm eff}$ values in KIC we estimated an error of $105 \; \mbox{K}$ for dwarfs and $165 \; \mbox{K}$ for giants. 
According to Table 8 and Figure 18 in \citet{kicteff}, the KIC $T_{\rm eff}$ values are approximately $200 \mbox{K}$ systematically too low for dwarfs and giants alike, so in reality the $T_{\rm eff}$ errors for the MARVELS target selection is likely closer to this value of 200~K.

This error on $T_{\rm eff}$ is larger than the 1-$\sigma$ error derived from benchmark grade stars given in \cite{casagrande10}. One reason is the assumed solar metallicity of our target stars, as noted above. A second reason is the fact that the Casagrande relations are calibrated for dwarfs and subgiants, but not for giants. For MARVELS target selection this is not important because we exclude all but the 6 brightest giants which are vetted manually using Vizier (see above). 
While $V - K_{S}$ as a temperature estimator is more sensitive to reddening than, e.g., $J - K_{S}$, it is less sensitive to metallicity errors. As reddening does not play a major role for the nearby dwarf stars that dominate the MARVELS targetsample, we consider $V - K_{S}$ to be a valid and optimal choice for our purposes (see Section \ref{sec:reddening} for an estimation of reddening and extinction).
In cases where the limits of the Casagrande relations for $J - K_{S}$ and $V - K_{S}$ allow us to compute a $T_{\rm eff}$ from both relations, we use the mean value.

\begin{figure}[htb]
\epsfig{file=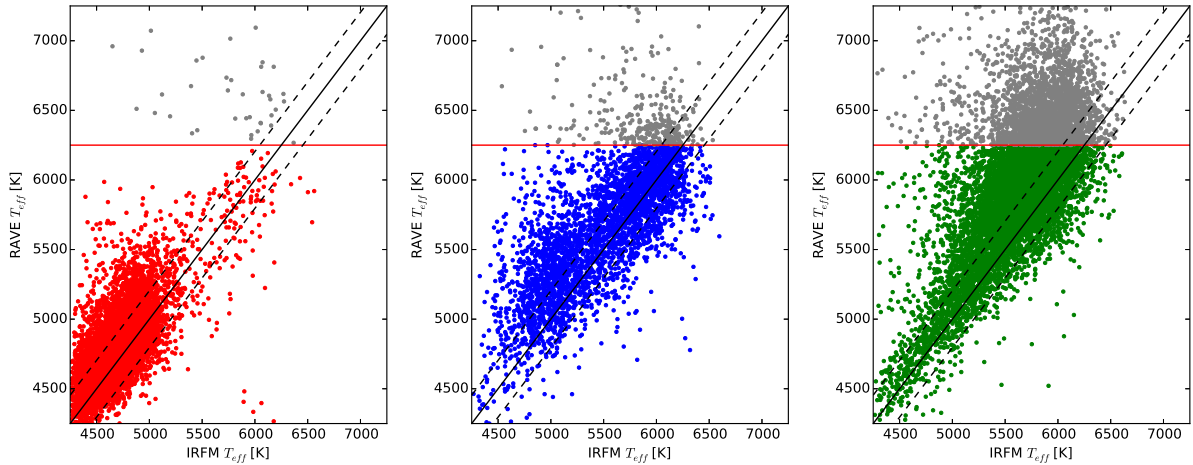, scale=0.43}
\centering
\parbox{16cm}{
\caption[Effective Temperatures from RAVE and IRFM] {\label{raveirfmteff} Effective temperature from RAVE and Infrared Flux Method. Solid black line: identity, dashed line: RAVE error ($204 \: \mbox{K}$), horizontal line: Marvels hot star cut-off at ($T_{\rm eff} > 6250$ K). Panels are by $\log(g)$ from RAVE: Left---giants; Center---subgiants; Right---Dwarfs}}
\end{figure}

We test the Casagrande relations with RAVE DR3, the result is shown in Figure \ref{raveirfmteff}. 
Running a least-squares fit over all stars we find an offset of $100 \pm 10 \; \mbox{K}$ between IRFM and RAVE temperatures. This offset matches the errors we estimate using KIC and the $85 \pm 14 \; \mbox{K}$ the RAVE team reports when comparing their data to high-resolution external results and is comparable to the $72 \pm 14 \; \mbox{K}$ RAVE reports for the general temperature offset in relation to external results. As reported by the RAVE team, the data show a wide spread of temperatures and a noticeable trend to yield higher temperatures especially for dwarf stars. However, the shift is about one MK-subclass---say G5 instead of G4---and not significant for the MARVELS target selection.

While the agreement between IRFM based temperatures and the RAVE spectroscopic temperatures is within the error margin for dwarfs with $T_{\rm eff} \le 5000 \; \mbox{K}$, it degrades quickly for hotter dwarfs.
Notably the agreement between RAVE and IRFM-derived temperatures for giants and subgiants is not worse than the agreement for hot dwarfs. Thus we conclude that the choice to use the Casagrande relations for giants and subgiants, although they are calibrated for dwarfs, is reasonable.

\subsection{Initial target selection process\label{sec:initial}}

\subsubsection{Basic input catalog construction}

Field names, center coordinates and number of observations for the year 1 and 2 fields are listed in table \ref{y12obs}. The basis for the target selection is the modified Guide Star Catalog 2.3 \citep{gsc23} as described in Section~\ref{sectinputcat}. 

Unlike in the final target selection, for the 1000 brightest stars matching the brightness and color cut of $J - K_{S} \ge 0.29$, a spectroscopic snapshot was taken by the SDSS double spectrograph, mainly used for SEGUE \citep[][]{segue}. The instrument has a resolution of $R \sim 2000$ and is described in greater detail in Section 2 of \citet{dspectro}. The double spectrograph saturates at $V = 9$, thus brighter stars needed special treatment during the initial phase. The stellar parameters $T_{\rm eff}$, $\log(g)$ and $[\rm Fe/ \rm H]$ were derived using a modified version of the SEGUE Stellar Parameter pipeline \citep[SSPP][]{sspp}. The target selection was a two-step process:

\begin{enumerate} \itemsep0pt
\item{Select up to 1000 stars for stellar characterization with SDSS spectrograph}
\item{Using the characterization from step 1 select 100 stars for drilling, 60 of them will get observed with MARVELS}
\end{enumerate}

For every star in a given field the steps in Table \ref{initial_preselsteps} were applied. The distance parameter in step 6 of the selection process ($62{''}$) is different from the $75{''}$ reported in step 5 of Table \ref{final_selsteps}.  The reason is that MARVELS switched to wider fibers in order to maximize the throughput in years 3 and 4. In addition the effective temperatures used here are derived from pre-observations using the modified SSPP pipeline instead of based on the Casagrande relations as in \ref{final_selsteps}. In order to allow a consistent comparison, we computed $T_{\rm eff}$ using the Casagrande relations for all stars tageted during the initial phase. 

\begin{table}[htb]
    \begin{tabular}{|c|p{7.5cm}|p{7.0cm}|}
        \hline
        Step & Criterion & Reason, Comments \\ \hline
          1  & Keep stars with $7.6 \le V \le 13.0$ & Include only stars in MARVELS magnitude limits \\
          2  & Keep $J - K_{S} \ge 0.29$    & Exclude stars that are clearly too hot \\
        3 & Ensure that positional coordinates, after correcting for proper motion, indicate that the star is in the field for at least 2 years from projected start of observations & Exclude stars that might wander off-plate \\
          4  & Closest star with $V < 9$ must be more than $5{''}$ away & Prevent flux contamination of target star \\
          5  & If two stars are closer than $62{''}$, keep the brighter star & Prefer bright stars for good SNR\\
          6  & Keep the brightest $1000$ stars & Limit the number of stars to number of SDSS spectrograph fibers \\
       \hline
          7  & stars must stay on plate for 2 years & Re-check because date of observation might have changed \\
          8  & Keep the 6 brightest giants & Only 6 giants wanted \\
          9  & Keep stars with $T_{\rm eff} < 6250 \; \mbox{K}$ & Exclude hot stars, $T_{\rm eff}$ from SSPP pipeline \\
         10  & Limit F stars (those with $5800 \mbox{K} \le T_{\rm eff} \le 6250 \mbox{K}$) to 40\% of all MARVELS targets in the field & Guarantee 50\% GK stars, $T_{\rm eff}$ from SSPP pipeline\\
         11  & If two stars are closer than $75{''}$, keep the brighter star & Prefer bright stars for good SNR \\
         12  & Limit the total number to 100 per field & 60 plugged, 40 as "reserve" in case of collision with guide stars \\
       \hline
    \end{tabular}
    \centering
    \parbox{16cm}{
        \caption[Initial pre-selection steps]{\label{initial_preselsteps} Steps for pre-selecting targets for spectrographic snapshot observations with the SDSS spectrograph (steps 1--6) and for observation with MARVELS (steps 7--12). $T_{\rm eff}$ from the modified SSPP pipeline.}}
\end{table}

The observations took place during twilight at the Apache Point Observatory. The spectra were evaluated for $T_{\rm eff}$, $\log(g)$ and $[\rm Fe/ \rm H]$ using the adapted SEGUE Stellar Parameter pipeline (SSPP). The stars were then split in a bright ($7.6 \le V \le 9.0$) and a faint ($9.0 < V < 13.0$) sample. Both samples were split between main sequence stars ($\log(g) >= 3.0$) and giants ($\log(g) < 3.0$). Although this split is different from the classification introduced with the final target selection ($\log(g) < 3.5$ for giants) it does not play a role in the large giant contamination of the initial phase. Only 10 stars from all stars observed in this phase have $3.0 \le \log{g} \le 3.5$ from the SSPP-pipeline and are flagged as \rpmj-giants, thus this shift is not responsible for the high contamination by giants in the initial phase.

The bright stars were checked against SIMBAD and usually rejected---allowing for special targets---if any of the following conditions were met:
\begin{enumerate} \itemsep0pt
\item{The spectral type was not between late F and early K for Main Sequence Stars or between mid G and early K for giants.}
\item{They are known variable stars.}
\item{They are in a visual binary with a companion less than $5{''}$ away.}
\item{They are known exoplanet hosts (except for benchmark stars).}
\item{Any anomalies were found making it unlikely that MARVELS could detect a substellar companion.}
\end{enumerate}

Bright stars passing these tests were combined with the faint star sample and steps 7 to 12 from Table \ref{initial_preselsteps} were applied. While it may appear that step 7---the first step after pre-selection---is redundant with step 3, several months may have elapsed between pre-selection and step 7. In this time the planned observations may have been delayed to a later date, thus necessitating a new check that the target stays on plate even with the new, later observation start date.


Although the spectrograph has only 64 fibers, we keep 100 stars in order to have a "reserve" if it turns out that a star can not be plugged because it is too close to a guiding star or for other technical reasons.

\subsubsection{Giant contamination in initial target selection}

\begin{figure}[htb]
\epsfig{file=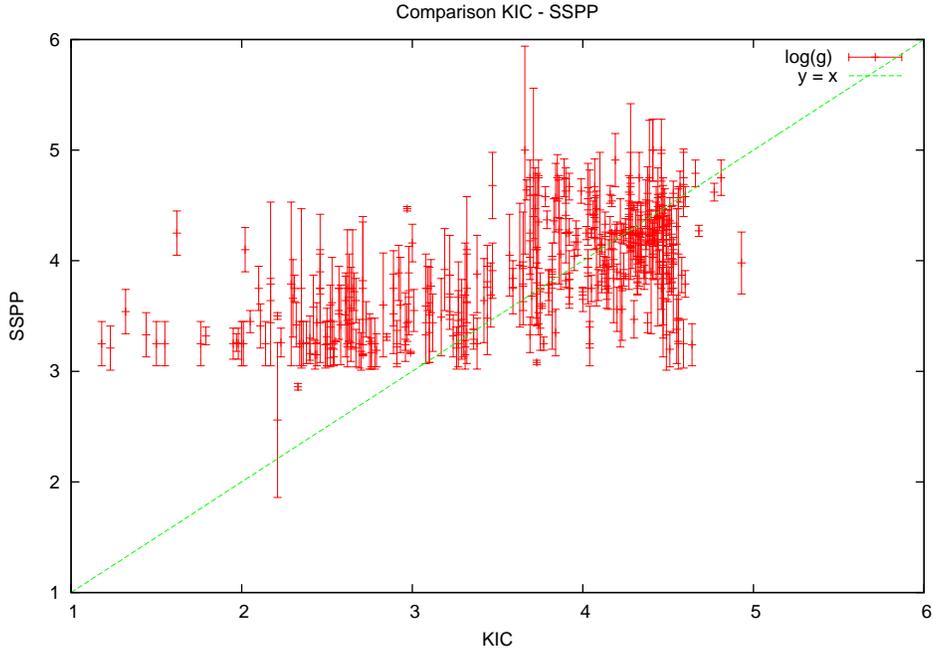, scale=1.0}
\centering
\parbox{16cm}{
\caption[$\log(g)$ with SDSS spectra and from KIC]{\label{loggfig} $\log(g)$ from spectra taken with SDSS and from KIC; the diagonal line represents identity.}}
\end{figure}

\begin{figure}[htb]
\epsfig{file=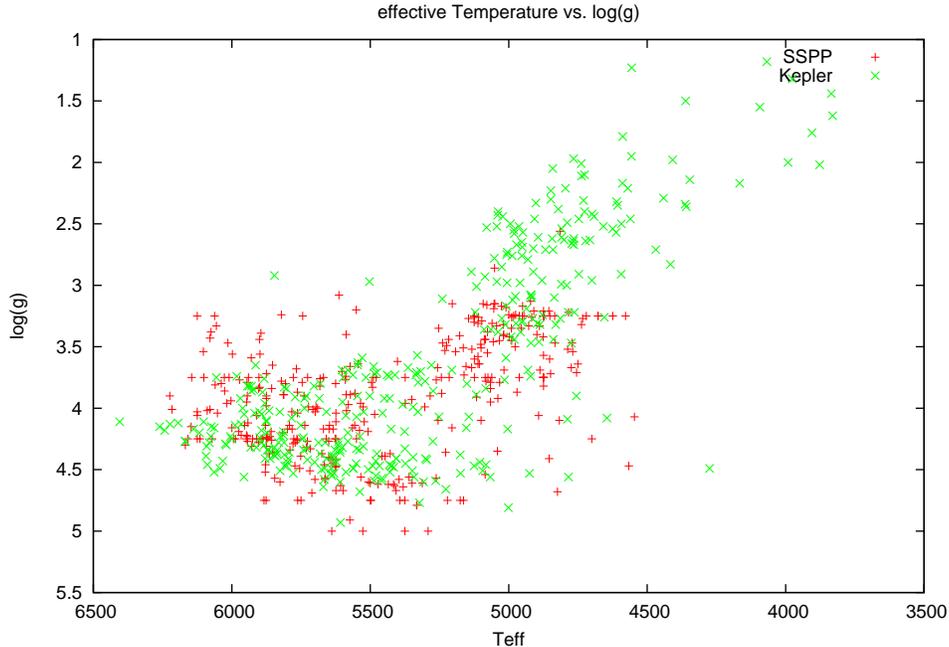, scale=1.0}
\centering
\parbox{16cm}{
\caption[$T_{\rm eff}$ vs. $\log(g)$ for SDSS and KIC]{\label{teffloggfig} $T_{\rm eff}$ vs. $\log(g)$ for the same stars from KIC and via a modified SSPP-pipeline for SDSS spectra. Both axis are inverted to resemble the display of an HR-diagram}}
\end{figure}

It was initially assumed that in this process $10 \%$ of the selected dwarfs would actually be giants due to errors in the $\log(g)$ determinations from the SSPP, yielding a final giant fraction of about $15 \%$ in the final sample. Instead, the contamination rate was about $35 \%$, as determined later by the \rpmj\ method. Some MARVELS fields overlapped with the Kepler field, and we compared the stellar characteristics obtained from SDSS spectra with those in the Kepler Input Catalog (KIC). Figure \ref{loggfig} shows the $\log(g)$ values from SDSS spectra versus values in the KIC. 
At $T_{\rm eff} < 5000 \; \mbox{K}$ results diverge rapidly and there is no agreement at all for $T_{\rm eff} < 4500 \; \mbox{K}$. Moreover, 
for all stars with KIC $\log(g) < 3$ the values disagree strongly. 
Given the fact that the original KIC values for $\log(g)$ are up to 1.0 dex too high \citep{casagrande14}, the true contamination by giants in the SSPP selected sample is even higher than suggested by Figure \ref{loggfig}.
Figure \ref{teffloggfig} shows the HR diagram (effective temperature and $\log(g)$) for the same set of stars from KIC and the SSPP-pipeline modified for MARVELS, again showing the strong discrepancies in the SSPP estimated $\log(g)$ values for cool giants.
We conclude that cool giants are misidentified as dwarfs by the modified SSPP pipeline. We therefore abandoned the spectroscopic pre-observations in favor of the streamlined target selection process described in section \ref{sectfinal}.

\section{Results\label{sec:results}}

\subsection{Summary of selected stars}

\begin{figure}[htb]
\epsfig{file=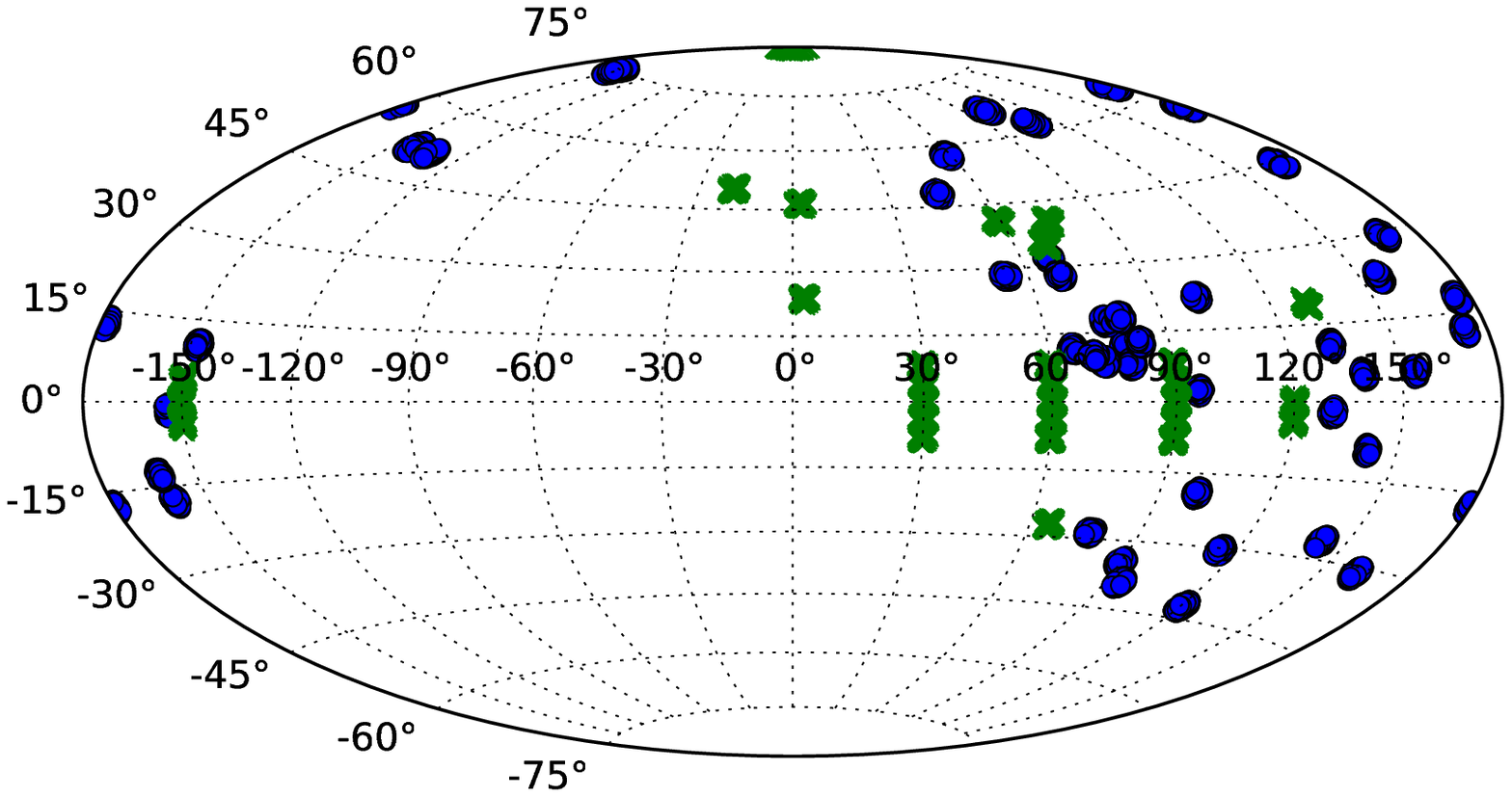, scale=0.75}
\centering
\parbox{16cm}{
\caption[MARVELS footprint]{\label{figfootprint} MARVELS footprint in galactic coordinates. Blue circles: target fields during the initial phase; green crosses: target fields during the final phase. The field centers along the galactic plane are located at latitudes of $-8$, $-4$, 0, 4 and 8 degrees and thus appear to blend into each other.}}
\end{figure}

The initial phase was significantly longer than the final phase---26 versus 15 months. The number of stars selected for observation reflects this asymmetry: 4130 stars in the initial phase, and 2900 stars for the final phase---adding to 7030 stars designed for observation out of which 5520 actually got observed. Figure \ref{figfootprint} shows the distribution on the sky in galactic coordinates.
The field centers along the galactic plane are from the final phase and located at galactic latitudes of $-8$, $-4$, 0, 4 and 8 degrees, and thus appear to blend into each other.

\begin{figure}[htb]
\epsfig{file=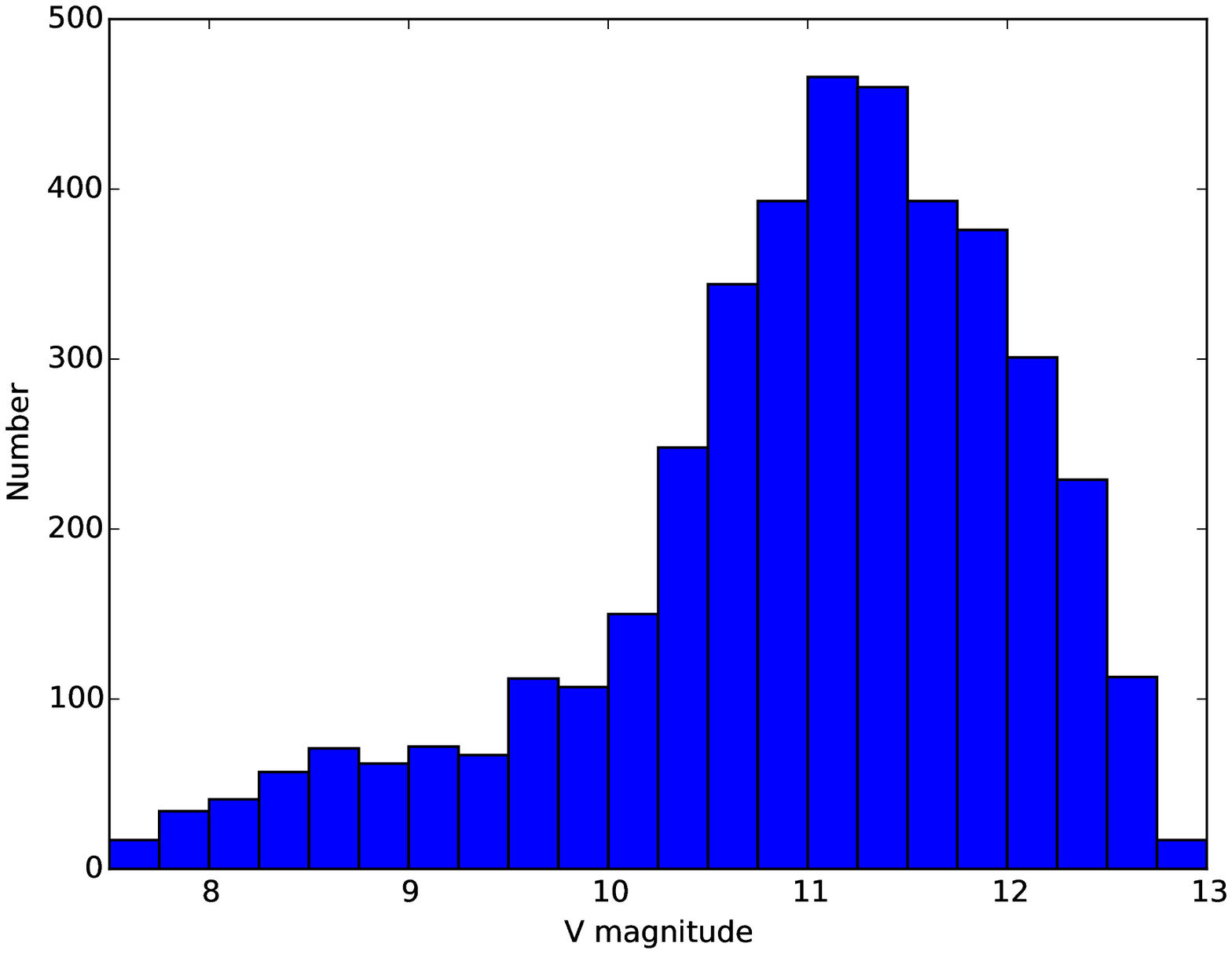, scale=0.40}
\epsfig{file=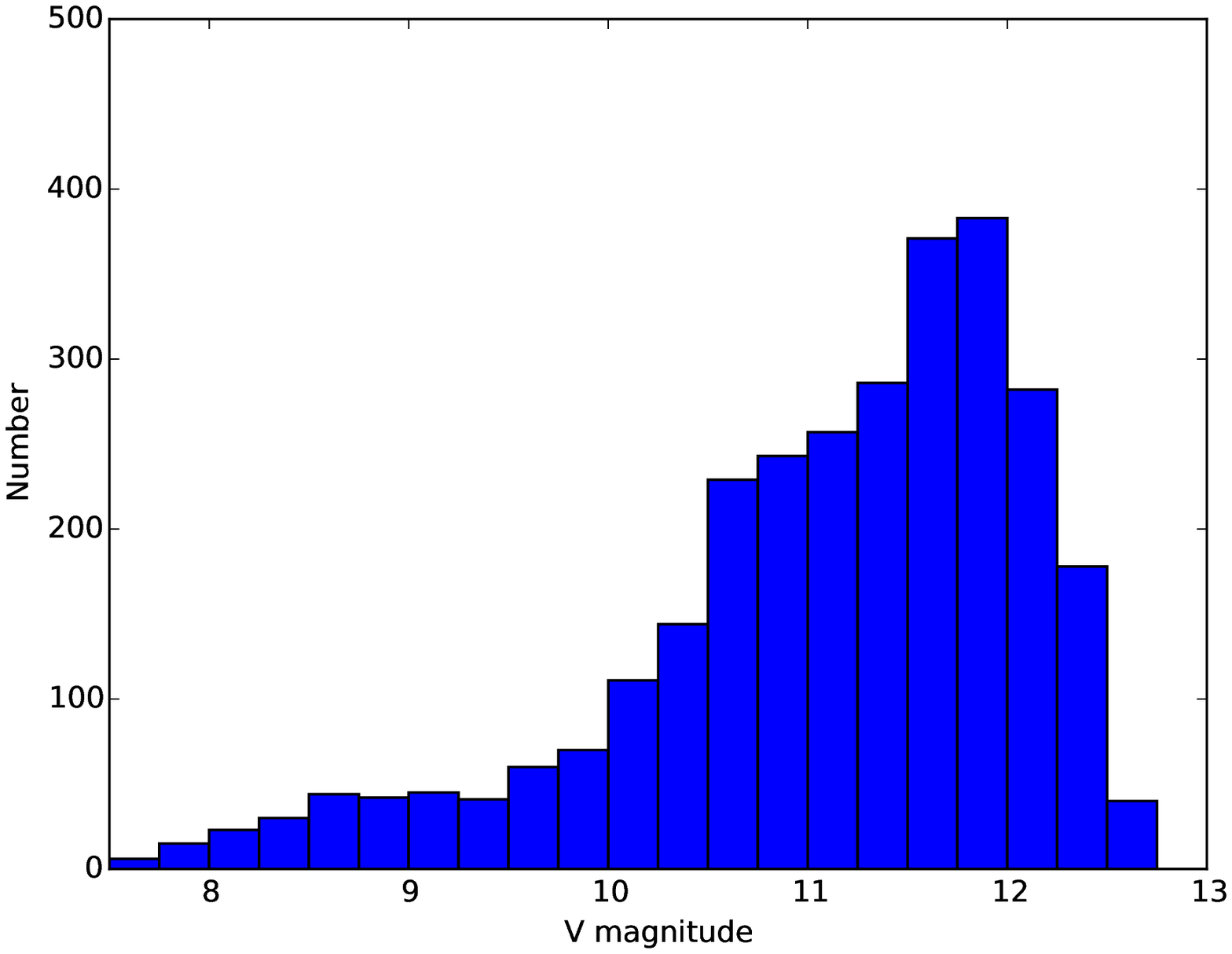, scale=0.40}
\centering
\parbox{16cm}{
\caption[Magnitude distribution]{\label{figvmagdist} V magnitude distribution for the initial phase (left) and final phase (right).}}
\end{figure}

Figure \ref{figvmagdist} shows the magnitude distribution in the V-band, for the initial phase at the left, for the final phase at the right. Aside from the different total numbers mentioned above, the most pronounced difference is a shift of the maximum by $0.5 \; \mbox{mag}$---from around $11.25 \; \mbox{mag}$ for the initial to $11.55 \; \mbox{mag}$ for the final phase. The reason is that coordination with APOGEE placed the fields outside of the galactic plane.  Since MARVELS and APOGEE could not observe the same stars in those sparse fields, the available stars were fainter. 

\begin{figure}[htb]
\epsfig{file=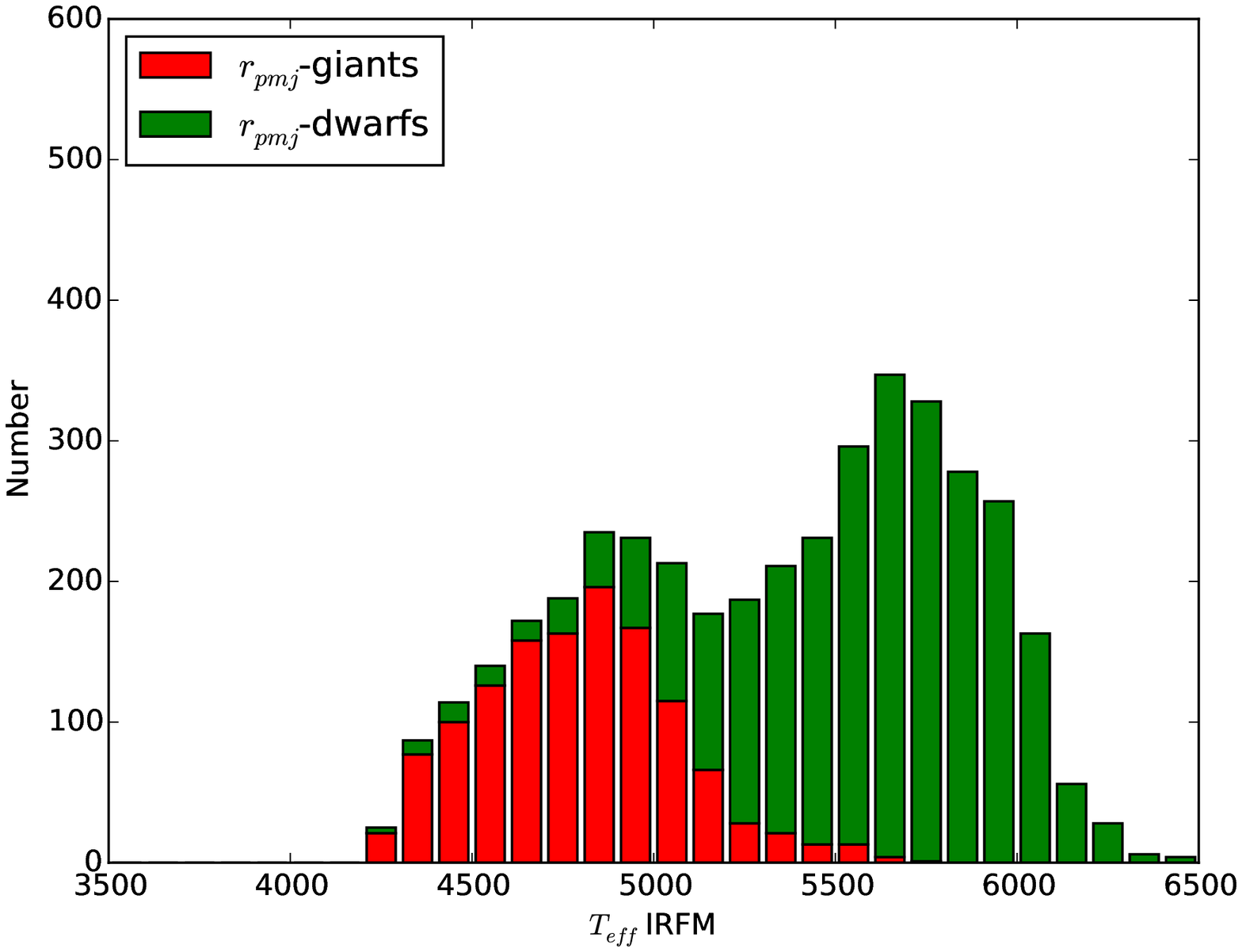, scale=0.40}
\epsfig{file=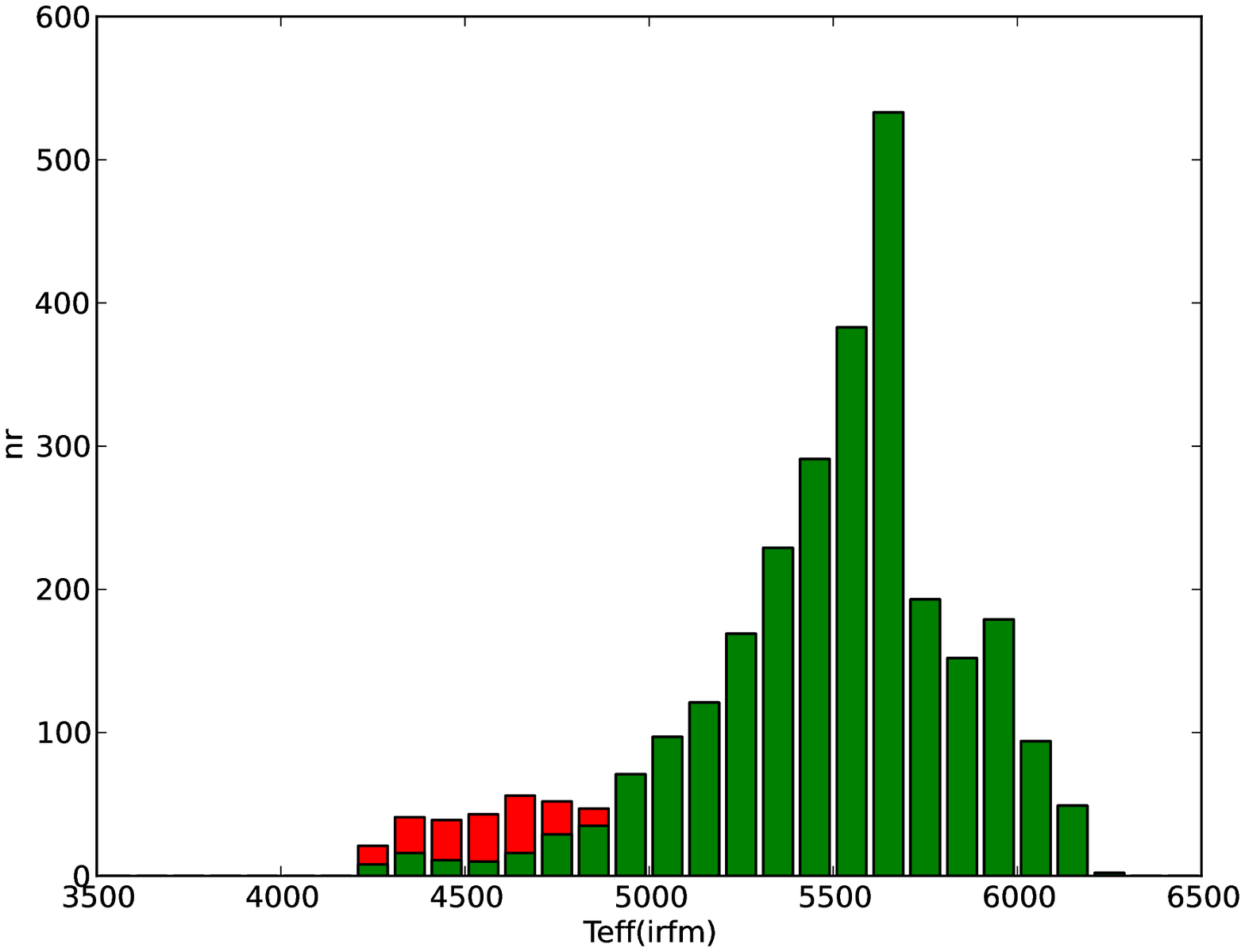, scale=0.40}
\centering
\parbox{16cm}{
\caption[Distribution of effective temperatures]{\label{figirfmteffstacked} Distribution of effective temperatures (IRFM method). Left: initial phase, right: final phase.}}
\end{figure}

Figure \ref{figirfmteffstacked} is a stacked histogram for effective temperatures of \rpmj-dwarfs and \rpmj-giants during the initial phase (left) and the final phase (right). For this comparison we computed effective temperatures for all stars from the initial phase using the Casagrande relations in order to allow a direct comparison to the temperatures estimated in the final phase. In the initial phase giants are overrepresented for $T_{\rm eff} < 5000 \; \mbox{K}$, indicating again that the $\log(g)$ values from the modified SSPP-pipeline were unreliable for cooler stars. Out of the 4130 stars selected for observation during the initial phase, 1414 stars are flagged as giants by the \rpmj\ method. This is $34 \%$ of the sample.

To estimate the fraction of giants in our sample we take the rates for MARVELS-selected RAVE stars from Table \ref{hip_rave}. About $14 \%$ of the \rpmj-giants are false positives, and thus are either dwarfs or sub-giants. On the other hand $3 \%$ of the \rpmj-dwarfs are false positives, and therefore giants. The estimated rate for the initial phase is then $34 - 0.14 \times 34 + 0.03 \times 66 = 31 \%$. For the final phase we manually checked 6 giants per field ($10 \%$). To this we add the $4 \%$ error for \rpmj-dwarfs and thus end with $14 \%$ giants in stars selected for the final phase. Given that we do want $10 \%$ of the stars to be giants, the difference between estimated and wanted giants gives the contamination rate. The results are summarized in Table \ref{tabgiantcontam}.

\begin{table}[htb]
    \begin{tabular}{|l|c|c|}
        \hline
        phase         & initial   & final    \\ \hline
        \rpmj-giants   &  34 \%    &  10 \%   \\
        \rpmj-dwarfs   &  66 \%    &  90 \%   \\
        est. giants   &  31 \%    &  14 \%   \\
        wanted        &  10 \%    &  10 \%   \\
        contamination &  21 \%    &   4 \%   \\
        \hline
    \end{tabular}
    \centering
    \parbox{6cm}{
        \caption[giant rates]{\label{tabgiantcontam} Giant comtamination rates for the initial and final phase.}}
\end{table}

\subsection{Effects of Reddening and Extinction on inferred stellar properties \label{sec:reddening}}

Most of the target fields of year 3 and 4 are located near the galactic plane ($-8 \le b \le 8$), so reddening and extinction might have to be taken into account. Using the \rpmj\,cut to distinguish dwarfs and subgiants from giants, extinction moves stars down in the \rpmj\,diagram, reddening moves them to the right. There are 3 possible effects:
\begin{itemize} \itemsep0pt
    \item Giants are pushed downwards over the cut by extinction, contaminating our sample.
    \item Dwarfs are shifted out of the region of interest by reddening and are lost.
    \item Hot stars are moved downward and to the right into our region of interest, polluting the sample.
\end{itemize}

We estimate the effect on a typical field for the first 2 years of observation (Kepler field) and---as worst case scenario---when observing directly towards the galactic center (assuming that redenning and extinction both increase towards the galactic center).

Taking the absolute magnitudes for dwarfs and giants from \citet{allen} and extending to bluer colors  using \citet{landolt82}, we computed the spectroscopic distances for dwarfs and giants of different spectral types with apparent magnitudes of $V = 10$ and $13$, representing the bright and faint end of our targets. We then computed the typical proper motions these stars would have according to the galactic model of \citet{slowpokes} and placed them into a \rpmj\ diagram.

\begin{figure}[htb]
\epsfig{file=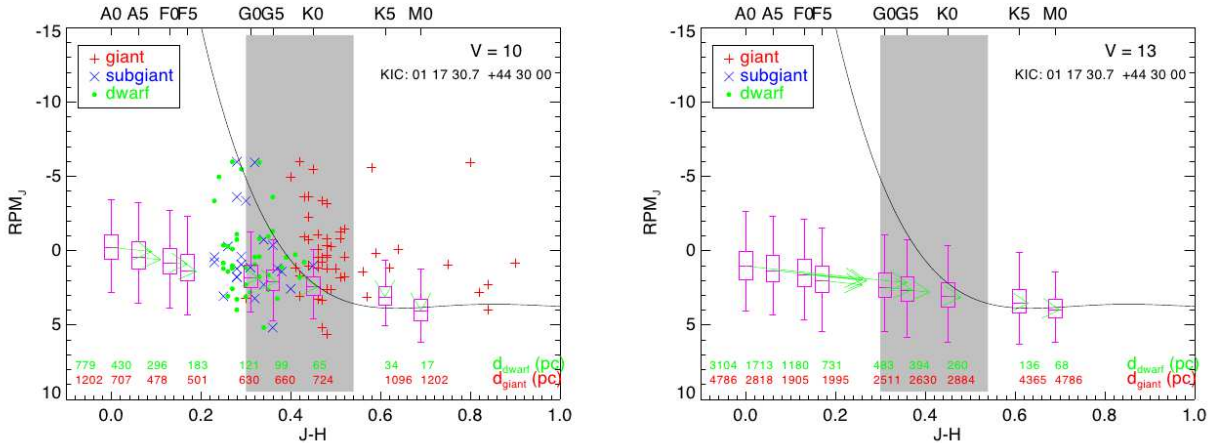, scale=0.490}
\centering
\parbox{16cm}{
\caption[Reddening and Extinction for the Kepler field] {\label{rpmjkic} Reddening and extinction for the Kepler field. Left panel: stars with $V = 10$; Right panel: stars with $V = 13$. Grey shade: MARVELS region of interest. Box and whiskers: galactic model position of stars with given magnitude and spectral type. Numbers on the lower legend are the typical distances of dwarfs (top, green) and giants (bottom, red)}}
\end{figure}

Figure \ref{rpmjkic} shows the result for the Kepler field with stars with $V = 10$ in the left and stars with $V = 13$ in the right panel. The MARVELS region of interest is shaded. For illustration we plotted the overlap stars of Kepler and MARVELS with colored symbols in the left panel. In order not to overcrowd the right panel showing the faint end of MARVELS magnitude range, we do not overplot the MARVELS-Kepler overlap stars, thus keeping the reddening/extiction vector more visible. The typical position of stars of a given spectral type and magnitude according to the galactic model from \citet{slowpokes} are marked as box and whiskers. Each box represents $50\%$ of all stars, the whiskers the upper and lower $27\%$ - leaving $3\%$ outliers apart. The spectroscopic distances of dwarfs and giants are given at the bottom. For each of the boxes we computed a reddening-extinction vector. We multiplied an assumed mean density of $N_H = 1 \; \mbox{atom/ccm}$ with the spectroscopic distance, yielding a column density of $n_H = N_H d$. This column density we converted to $A_V$ and further to  $E(B-V)$, adopting the relations

\begin{equation}
    A_V = n_H / 2.30 \times 10^{21}, \quad  E(B-V) = A_V / 3.1
\end{equation}

Taking $A_J / A_V = 0.282$ and $A_H / A_V = 0.190$ from \cite{cardelli89} we converted $E(B-V)$ to $E(J-H)$ which completes the vector $(E(J-H), A_{J})$. For dwarfs we plotted this vector at each box. For $V = 10$ there is no noticeable shift in and out of the region of interest. For $V = 13$ the brightest stars (A0) can be shifted in the region of interest. However, they are a very small fraction of the stellar population and thus will not significantly pollute the sample.

\begin{figure}[htb]
\epsfig{file=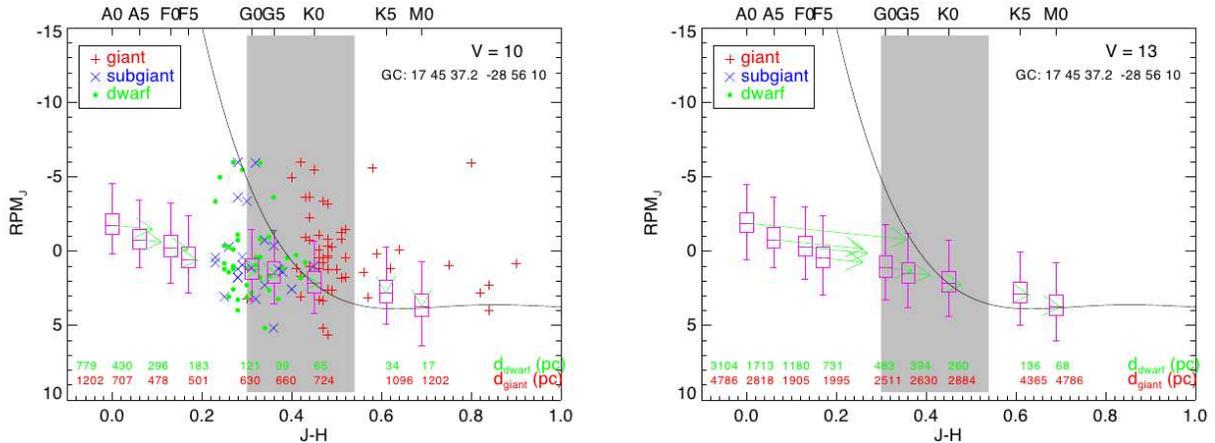, scale=0.490}
\centering
\parbox{16cm}{
\caption[Reddening and Extinction for the Galactic Center] {\label{rpmjgc} Reddening and extinction for the Galactic Center. Left panel: stars with $V = 10$; Right panel: stars with $V = 13$. Grey shade: MARVELS region of interest. Box and whiskers: galactic model position of stars with given magnitude and spectral type. Numbers on the lower legend are the typical distances of dwarfs (top, green) and giants (bottom, red)}}
\end{figure}

We repeated the same analysis for observations towards the Galactic Center, with the results shown in figure \ref{rpmjgc}. Compared to the Kepler field there is a slight but insignificant shift in the position of the boxes. For bright stars ($V = 10$, left panel) reddening and extinction do not play a significant role. For faint stars ($V = 13$, right panel) A0 stars get shifted into the region of interest. We might see pollution by late F-dwarfs; A5 to F5 are not reddened enough. As even in the worst case of observing towards the Galactic Center we only would see a light pollution by late F-dwarfs, we concluded that correcting for reddening and extinction is not necessary for the MARVELS target selection.

Reddening and extinction get stronger if we assume a higher density than average 1 atom per ccm. The spectroscopic distance of giants with an apparent magnitude of $V = 13$ is $4.8 \; \mbox{kpc}$. As long as we do not hit a denser region within this distance, our estimation of reddening and extinction holds.

\section{Summary\label{sec:summary}}
In this paper we have discussed the targe selection methodology for creating an input catalog for the MARVELS radial velocity survey. The MARVELS survey was interested in looking for radial velocity companions to stars of the FGK spectral type primarily focusing on dwarf stars. To achieve this goal, a target selection criteria of 10\% giant stars 90\% dwarf stars was set. Target selection for MARVELS is broken down into two distinct phases, the initial phase which found targets for the first half of the survey, and the final phase which found targets for the second half of the survey. The initial target selection method used low-resolution ($R \sim 2000$) spectra from the SDSS spectrographs processed by a modified SSPP pipeline. This method ulitmately proved to be inadequate for removing giant star contamination primarly on the cool end ($T_{\rm eff} < 4500$). This result was not entirely unexpected because the MARVELS stars are significantly brighter than what the SSPP was designed to work with. As a result, the giant contamination rate for the initial phase was 31\%. Given the results of the initial phase, the final phase of target selection used a different method. Instead of low-resolution spectra, a reduced proper motion method (\rpmj) was employed. This method did a much better job meeting our criteria providing a giant contamination rate of just 13\%. 

This investigation also revealed two other notable results. First is that interstellar reddening is not a major factor in influencing the stars selected for MARVELS. This is due to primarily to the relatively short distances to the MARVELS stars and the way in which the reddening vector points in the \rpmj~diagram. Second is that the \rpmj~method, although being quite useful for separating dwarfs and giants, is not able to adequately separate dwarfs and subgiants (subgiant contamination of the dwarf sample is on order 30\%). For the MARVELS scientific goals this was not important. However, future surveys or missions need to be aware of this fact when designing their respective input catalogs.

\acknowledgments

We acknowledge the generous support from the W.M. Keck Foundation for developing the MARVELS survey instruments. The MARVELS project was also supported from NSF with grant AST-0705139, NASA with grant NNX07AP14G and the University of Florida. 

Funding for SDSS-III has been provided by the Alfred P. Sloan Foundation, the Participating Institutions, the National Science Foundation, and the U.S. Department of Energy Office of Science. The SDSS-III web site is \url{http://www.sdss3.org/}.

SDSS-III is managed by the Astrophysical Research Consortium for the Participating Institutions of the SDSS-III Collaboration including the University of Arizona, the Brazilian Participation Group, Brookhaven National Laboratory, Carnegie Mellon University, University of Florida, the French Participation Group, the German Participation Group, Harvard University, the Instituto de Astrofisica de Canarias, the Michigan State/Notre Dame/JINA Participation Group, Johns Hopkins University, Lawrence Berkeley National Laboratory, Max Planck Institute for Astrophysics, Max Planck Institute for Extraterrestrial Physics, New Mexico State University, New York University, Ohio State University, Pennsylvania State University, University of Portsmouth, Princeton University, the Spanish Participation Group, University of Tokyo, University of Utah, Vanderbilt University, University of Virginia, University of Washington, and Yale University.


\bibliography{myrefs}

\clearpage

\appendix
\section{Observed Fields}
Table \ref{y12obs} provides fieldnames, coordinates and number of observations during the initial phase,
Table \ref{y34obs} provides the same information for the final phase.
\begin{table}
    \caption{ Name, center coordinates and number of observations
    for fields from the initial phase}
    {\small
    \begin{tabular}{|l|d{2}|d{2}|d{5}|d{5}|c|}
        \hline
        Name        & RA       & Dec     & l           & b          &  Obs \\
        \hline
        47UMA       & 164.86   & 40.43   & 175.787     & 63.3645    & 29 \\
        51PEG       & 344.37   & 20.77   & 90.06669    & -34.72791  & 36 \\
        FIELD1068   & 264.3    & 30      & 54.18431    & 28.24962   & 23 \\
        FIELD1110   & 100.8    & 33      & 182.19045   & 12.7985    & 23 \\
        FIELD1348   & 262.86   & 42      & 67.41378    & 32.0944    & 23 \\
        FIELD1349   & 268.57   & 42      & 68.33689    & 27.92775   & 25 \\
        FIELD1572   & 321.7    & 54      & 95.6606     & 2.33948    & 31 \\
        FIELD1631   & 101.54   & 60      & 155.71982   & 22.6218    & 34 \\
        GJ176       & 70.73    & 18.96   & 180.01238   & -17.42718  & 30 \\
        GJ436       & 175.55   & 26.71   & 210.53025   & 74.57257   & 29 \\
        GL273       & 111.85   & 5.23    & 212.33978   & 10.37355   & 28 \\
        HAT-P-1     & 344.45   & 38.67   & 99.7928     & -19.04242  & 36 \\
        HAT-P-3     & 206.09   & 47.97   & 100.09623   & 66.74412   & 24 \\
        HAT-P-4     & 229.99   & 36.13   & 58.29627    & 57.34666   & 6 \\
        HD118203    & 203.51   & 53.73   & 109.3461    & 62.26017   & 26 \\
        HD17092     & 41.59    & 49.65   & 141.3032    & -9.07813   & 30 \\
        HD17156     & 42.44    & 71.75   & 131.99263   & 10.99141   & 38 \\
        HD219828    & 349.69   & 18.65   & 94.26312    & -39.00768  & 36 \\
        HD37605     & 85.01    & 6.06    & 199.10635   & -12.89319  & 28 \\
        HD4203      & 11.17    & 20.45   & 120.78828   & -42.39362  & 36 \\
        HD43691     & 94.89    & 41.09   & 172.65004   & 11.92868   & 31 \\
        HD46375     & 98.3     & 5.4     & 206.04629   & -1.57714   & 25 \\
        HD49674     & 102.88   & 40.87   & 175.33733   & 17.37424   & 26 \\
        HD68988     & 124.59   & 61.46   & 155.26411   & 33.91797   & 40 \\
        HD80606     & 140.66   & 50.54   & 167.51619   & 44.3251    & 29 \\
        HD88133     & 152.53   & 18.12   & 217.91036   & 51.87233   & 25 \\
        HD89307     & 154.59   & 12.56   & 227.39155   & 51.36809   & 29 \\
        HD89744     & 155.54   & 41.17   & 178.49592   & 56.40393   & 35 \\
        HD9407      & 23.64    & 68.95   & 126.80841   & 6.40343    & 29 \\
        HIP14810    & 47.81    & 21.1    & 161.54971   & -31.09206  & 31 \\
        K10         & 294.12   & 46.01   & 78.80125    & 11.96774   & 23 \\
        K14         & 299.64   & 44.87   & 79.67739    & 8.02455    & 23 \\
        K15         & 296.12   & 43.53   & 77.23814    & 9.55789    & 20 \\
        K20         & 294.71   & 39.63   & 73.25579    & 8.6313     & 23 \\
        K21         & 291.58   & 38.15   & 70.78675    & 10.11117   & 19 \\
        K4          & 295.69   & 49.9    & 82.83844    & 12.79648   & 19 \\
        K5          & 291.93   & 48.45   & 80.38966    & 14.37021   & 20 \\
        K7          & 285.05   & 45.2    & 75.36922    & 17.43255   & 20 \\
        K8          & 281.91   & 43.44   & 72.80583    & 18.91956   & 26 \\
        KEPLER3-TRES2 & 285.9   & 49.2   & 79.51202   & 18.32046   & 21 \\
        KEPLER4     & 282.52   & 47.46   & 76.97688    & 19.84864   & 23 \\
        WASP-1      & 5.17     & 31.99   & 115.36902   & -30.42966  & 36 \\
        XO-1        & 240.55   & 28.09   & 45.73607    & 48.00637   & 20 \\
        XO-2        & 117.03   & 50.16   & 168.36412   & 29.32105   & 31 \\
    \hline
    \end{tabular}
    }
    \label{y12obs}
\end{table}

\begin{table}
    \caption{\label{y34obs} Name, center coordinates and number of observations
    for fields from the final phase}
    {\small
    \begin{tabular}{|l|d{5}|d{5}|d{2}|d{2}|c|}
        \hline
        Name         & RA         & Dec       & l      & b       & Obs \\
        \hline
        APG\_030-04   & 285.09054  & -4.42875  & 30     & -4      & 2 \\
        APG\_030-08   & 288.67708  & -6.23117  & 30     & -8      & 6 \\
        APG\_030+00   & 281.5215   & -2.60914  & 30     & 0       & 2 \\
        APG\_030+04   & 277.96279  & -0.77944  & 30     & 4       & 1 \\
        APG\_060-04   & 299.71129  & 21.85097  & 60     & -4      & 3 \\
        APG\_060+00   & 295.97625  & 23.89036  & 60     & 0       & 2 \\
        APG\_060+04   & 292.12433  & 25.83681  & 60     & 4       & 3 \\
        APG\_060+08   & 288.14717  & 27.67833  & 60     & 8       & 1 \\
        APG\_090-04   & 322.15829  & 45.5085   & 90     & -4      & 3 \\
        APG\_090-08   & 325.90667  & 42.55142  & 90     & -8      & 1 \\
        APG\_090+04   & 313.379    & 50.984    & 90     & 4       & 2 \\
        APG\_090+08   & 308.21592  & 53.43419  & 90     & 8       & 1 \\
        APG\_120-08   & 7.82354    & 54.757    & 120    & -8      & 3 \\
        APG\_120+04   & 5.44675    & 66.70278  & 120    & 4       & 4 \\
        APG\_120+08   & 4.05675    & 70.67103  & 120    & 8       & 3 \\
        APG\_150-04   & 57.03879   & 49.35553  & 150    & -4      & 4 \\
        APG\_150-08   & 53.44846   & 46.16447  & 150    & -8      & 4 \\
        APG\_150+04   & 65.78704   & 55.32392  & 150    & 4       & 3 \\
        APG\_150+08   & 71.16704   & 58.02211  & 150    & 8       & 7 \\
        APG\_165-04   & 68.18012   & 36.21319  & 165    & -4      & 3 \\
        APG\_165+04   & 85.19104   & 45.86597  & 165    & 4       & 2 \\
        APG\_180-08   & 78.90013   & 24.56125  & 180    & -8      & 3 \\
        APG\_180+00   & 86.40483   & 28.93617  & 180    & 0       & 5 \\
        APG\_180+04   & 90.38625   & 30.96164  & 180    & 4       & 5 \\
        APG\_180+08   & 94.53463   & 32.85992  & 180    & 8       & 5 \\
        APG\_195-04   & 87.26887   & 11.99861  & 195    & -4      & 8 \\
        APG\_195+04   & 101.93729  & 19.56097  & 195    & 4       & 7 \\
        APG\_210-08   & 94.40717   & -1.05322  & 210    & -8      & 5 \\
        APG\_210+00   & 101.5215   & 2.60914   & 210    & 0       & 1 \\
        APG\_210+08   & 108.67708  & 6.23117   & 210    & 8       & 12 \\
        APG\_M13      & 250.51667  & 36.50722  & 59.01  & 40.91   & 24 \\
        APG\_M15      & 322.49292  & 12.16694  & 65.01  & -27.31  & 4 \\
        APG\_M53      & 198.23042  & 17.16917  & 330.53 & 78.87   & 24 \\
        APG\_M5PAL5   & 229.225    & 1.065     & 2.33   & 46.47   & 6 \\
        APG\_M92      & 259.28042  & 43.13639  & 68.33  & 34.85   & 3 \\
        APG\_N2420    & 114.55583  & 21.53333  & 198.13 & 19.58   & 14 \\
        APG\_N5634SGR2 & 216.375   & -5.66064  & 341.17 & 50.1    & 6 \\
        APG\_N6229    & 251.94167  & 47.5425   & 73.63  & 40.3    & 4 \\
        APG\_NGP      & 192.85958  & 27.12833  & 53.19  & 89.99   & 2 \\
        APG\_SGR1     & 123.76125  & 31.90489  & 190.2  & 30.7    & 21 \\
        APG\_VOD3     & 191        & -7.8      & 299.71 & 55.02   & 13 \\
        APGS\_M107    & 247.38125  & -13.05653 & 2.88   & 23.58   & 15 \\
        APGS\_M3      & 206.29667  & 27.62556  & 38.4   & 78.06   & 26 \\
        APGS\_N4147   & 181.52583  & 18.54194  & 250.02 & 76.49   & 24 \\
        APGS\_N5466   & 211.58417  & 28.7785   & 42.99  & 73.39   & 27 \\
        APGS\_VOD1    & 176.5      & 0.5       & 269.69 & 59.08   & 16 \\
        APGS\_VOD2    & 185.5      & -0.25     & 287.24 & 61.72   & 12 \\
        HD4203        & 11.172     & 20.38789  & 120.79 & -42.46  & 4 \\
        HD46375       & 98.3       & 5.4       & 206.05 & -1.58   & 10 \\
        HIP14810      & 47.809     & 21.01293  & 161.61 & -31.16  & 4 \\
        \hline
    \end{tabular}
    }
\end{table}

\end{document}